\documentclass[aps,prl,twocolumn,groupedaddress,showpacs]{revtex4}

\usepackage{graphicx}
\usepackage{dcolumn}
\usepackage{bm}
\usepackage{float}

\def\no{\noindent}
\def\bc{\begin{center}}
\def\ec{\end{center}}

\def\beq{\begin{equation}}
\def\eeq{\end{equation}}

\title{
Anderson localization in a correlated fermionic mixture
}
\begin{document}


\title{Anderson localization in correlated fermionic mixtures}

\author{O. Fialko and K. Ziegler}%
\email{Klaus.Ziegler@Physik.Uni-Augsburg.de}
\affiliation{
Institut f\"ur Physik, Universit\"at Augsburg, D-86135 Augsburg, Germany
}

\date{\today}

\begin{abstract}
A mixture of two fermionic species with different masses is studied in an optical lattice.
The heavy fermions are subject only to thermal fluctuations, 
the light fermions also to quantum fluctuations. We derive the Ising-like distribution 
for the heavy atoms and study the localization properties of the light fermions
numerically by a transfer-matrix method. In a two-dimensional system one-parameter 
scaling of the localization length is found with a transition from delocalized states 
at low temperatures to localized states at high temperature. The critical exponent 
of the localization length is $\nu\approx 0.88$.  
\end{abstract}

\maketitle



The question of Anderson localization in an ultra cold gas has attracted considerable 
attention recently
by a number of experimental groups \cite{ospelkaus06,roati08,billy08}. 
Although the phenomenon itself has been studied
in great detail over the last 50 years by many theoretical groups for various physical 
systems \cite{anderson58,abraham79,wegner76}, its experimental observation has been difficult. 
One of the reasons is that
Anderson localization is an interference effect of waves due to elastic scattering
in a random environment (disorder). Real systems, however, experience also substantial 
inelastic scattering (e.g. absorption of electromagnetic waves by the scattering atoms, 
Coulomb interaction in electronic systems etc.) This may hamper the direct observation
of Anderson localization significantly.
Another reason is that random scattering
is difficult to control in a real system. This is important in order to distinguish
Anderson localization from simple trapping due to local potentials. It requires some 
kind of averaging over an ensemble of randomly distributed scatterers.

Ultra cold gases
offer conditions, where most physical parameters are controllable. Since the atoms
are neutral, there is no Coulomb interaction, and at sufficiently high dilution the
interatomic collisions are negligible. Moreover, a periodic potential (optical lattice)
can be applied by counterpropagating laser fields. This enables us to control the 
kinetic properties of the gas atoms by creating a specifically designed dispersion.  
Disorder could be created by disturbing the periodicity of the optical lattice. 
In practice, however, this is not easy because real disorder would require infinitely
many laser frequencies. A first attempt is to study the superposition of two
laser fields with ``incommensurate'' frequencies (i.e. the ratio of the two frequencies
is an irrational number) \cite{roati08}. An alternative is to randomize the laser field
by sending it through a diffusing plate \cite{billy08}. 

Recent progress in atomic mixtures \cite{stan04} has offered another possibility to create disorder
in an atomic system. Mixing of two different atomic species, where one is heavier than the 
other, creates a situation where the light atoms are scattered by the randomly
distributed heavy atoms \cite{gavish05,ates05,maska08,ziegler08}. An optical lattice is applied
in order to keep the heavy atoms in quenched positions. Due to their higher mass,
the heavy atoms behave classically in contrast to the light atoms, which can
tunnel in the optical lattice. A crucial question is what determines the distribution
of the heavy atoms. The most direct distribution is obtained by putting atoms randomly
in the optical lattice ``by hand'', each of them with independent probability \cite{gavish05}.
This case corresponds to uncorrelated disorder. Another possibility is to fill the optical 
lattice with both atomic species and consider a repulsive (local) interaction 
between them. Then the two species have to arrange each other such that the total atomic
system presents a grand-canonical ensemble at a given temperature and a given lattice filling. 
In the presence of interparticle interaction within each atomic species there is a complex 
interplay of interaction and localization effect. This makes it difficult to isolate the 
effect of Anderson localization. In order to avoid interaction within each species we
choose spin-polarized fermions in an optical lattice. Then only
the Pauli principle controls the short-range interaction within each species,
such that the remaining interaction is between the different fermionic species.
It has been shown that then the light
atoms are subject to a quenched average with respect to a thermal distribution of the
heavy atoms, and that the distribution is related to an Ising-like
model \cite{ates05,ziegler06,maska08}. The latter implies (strong) correlations between the heavy
atoms. For systems in more than one dimension there is a critical temperature $T_c$ at which
the correlation length diverges. This system provides several interesting features for studying
Anderson localization. Although it is a many-body system, the light atoms behave
effectively like independent (spinless fermionic) quantum particles in a random potential. 
The correlation of the randomness can be controlled by temperature, where the correlation 
length decreases with increasing temperature for temperatures $T>T_c$, or by the strength
of the inter-species scattering.

In the following we shall study diffusion and Anderson localization in the grand-canonical 
ensemble of two spin-polarized fermionic species in one and two dimensions. 
Motivated by recent experimental study on a dilute BEC in $d=1$ \cite{billy08}, we
consider a realistic scenario, in which an initial state is prepared at the center 
with a trapping potential and then it is released by opening (i.e. switching off) the trap. 
The expansion of the wave function is observed within our numerical procedure.


{\it model:} $c^\dagger$ ($c$) are creation (annihilation) operators of the
light fermionic atoms, $f^\dagger$ ($f$) are the corresponding
operators of the heavy fermionic atoms. This gives the formal mapping
$
^6{\rm Li}\rightarrow c^\dagger_r, c_r$ and 
$
^{40}{\rm K}\ (^{23}{\rm Na},\ ^{87}{\rm Rb})
\rightarrow f^\dagger_r, f_r 
$.
\no
The physics of the mixture of atoms is defined by the asymmetric 
Hubbard Hamiltonian
\beq
H= -{\bar t}_c\sum_{\langle r,r'\rangle}c_r^\dagger c_{r'}  
-{\bar t}_f\sum_{\langle r,r'\rangle}f_r^\dagger f_{r'}
\]
\[ 
- \sum_r\Big[\mu_c c_{r}^\dagger c_{r}+\mu_ff_{r}^\dagger f_{r}
-Uf^\dagger_{r}f_{r}c^\dagger_{r}c_{r}\Big] .
\label{hamilton}
\eeq
The effective interaction within each species is controlled by 
the (repulsive) Pauli principle, whereas the interaction strength of
different atoms is $U$.
If the $f$ atoms are heavy, the related tunneling rate is very small. 
The limit ${\bar t}_f=0$ is known as the Falicov-Kimball model 
\cite{falicov69,farkasovsky97,freericks03,maska08}.

A grand-canonical ensemble of fermions at the inverse temperature 
$\beta=1/k_B T$ is defined by the partition function
\[
Z={\rm Tr} e^{-\beta H} .
\]


In the FK limit ${\bar t}_f=0$ the Hamiltonian of the light atoms
$H_c(\{ n_r\})$
depends only on the real numbers $\{ n_r\}$ ($n_r=0,1$), and it is a quadratic
form with respect to the $c$ operators of the light atoms:
\beq
\sum_{r,r'}h_{c;rr'}c^\dagger_{r}c_{r'}
= -{\bar t}_c\sum_{\langle r,r'\rangle}c_r^\dagger c_{r'} 
+ \sum_r (Un_r -\mu_c )c^\dagger_{r}c_{r}  \ .
\label{hamilton2}
\eeq
This means that the density fluctuations $n_r=f_r^\dagger f_r$ have been 
replaced by classical variables $n_r=0,1$.
Thus $H_c(\{ n_r\})$ describes
non-interacting fermions which are scattered by heavy atoms, represented by $n_r$. 
The trace ${\rm Tr}_c$ in the partition function can be 
evaluated and gives a fermion determinant:
\[
Z 
=\sum_{\{n_r\}}e^{\beta\mu_f\sum_r n_r}{\rm Tr}_c
\left(e^{-\beta H_c(\{ n_r\})}\right)
\]
\beq
=\sum_{\{n_r\}}e^{\beta\mu_f\sum_r n_r}
{\rm det}[{\bf 1}+e^{-\beta h_c}] \ .
\label{partition}
\eeq
The right-hand side is a sum over (non-negative)
statistical weights. After normalization we can define
\beq
P(\{ n_r\})={1\over Z}e^{\beta\mu_f\sum_rn_r}
{\rm det}[{\bf 1}+e^{-\beta h_c}]
\label{distr0}
\eeq
which gives $\sum_{\{n_r\}}P(\{ n_r\})=1$.
Thus $P(\{ n_r\})$ is a probability distribution for correlated disorder and
describes the distribution of the heavy atoms. In the strong-coupling regime 
${\bar t}_c^2/2U\gg 1$ the distribution becomes an Ising model with nearest-neighbor 
coupling. At half-filling (i.e. $\mu_f=\mu_c=U/2$) this reads \cite{ates05}
\beq
P(\{ S_r\})\propto \exp\left(-\beta({\bar t}_c^2/2U)\sum_{<r,r'>}S_r S_{r'}\right)
\label{idistr}
\eeq
where $S_r=2n_r-1$.


\begin{figure}[b]
\begin{center}$
\begin{array}{cc}
\includegraphics[width=4cm]{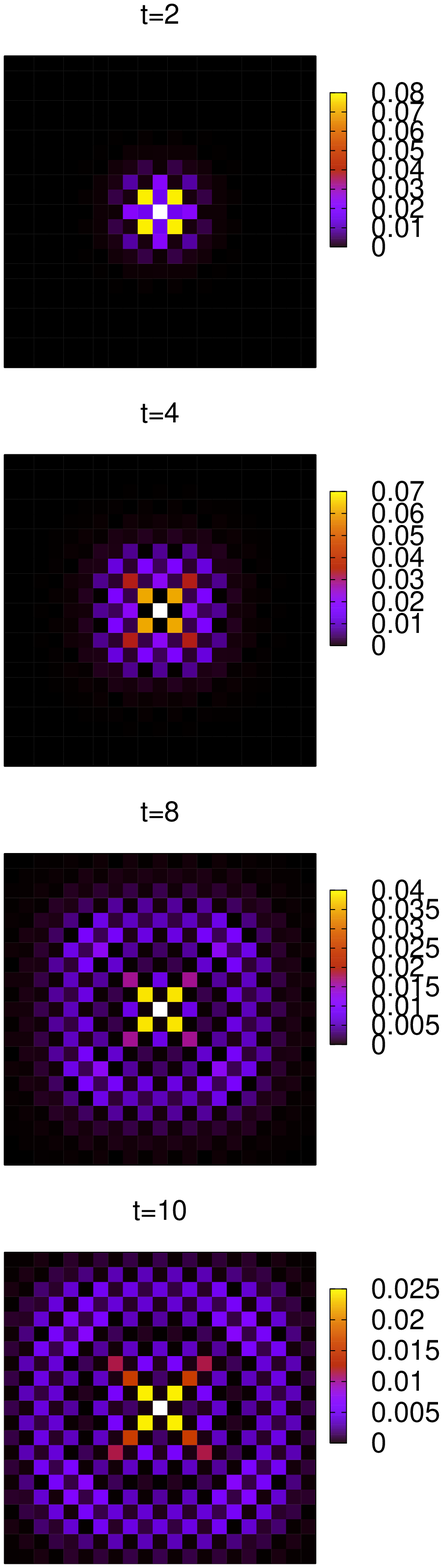} &
\includegraphics[width=4cm]{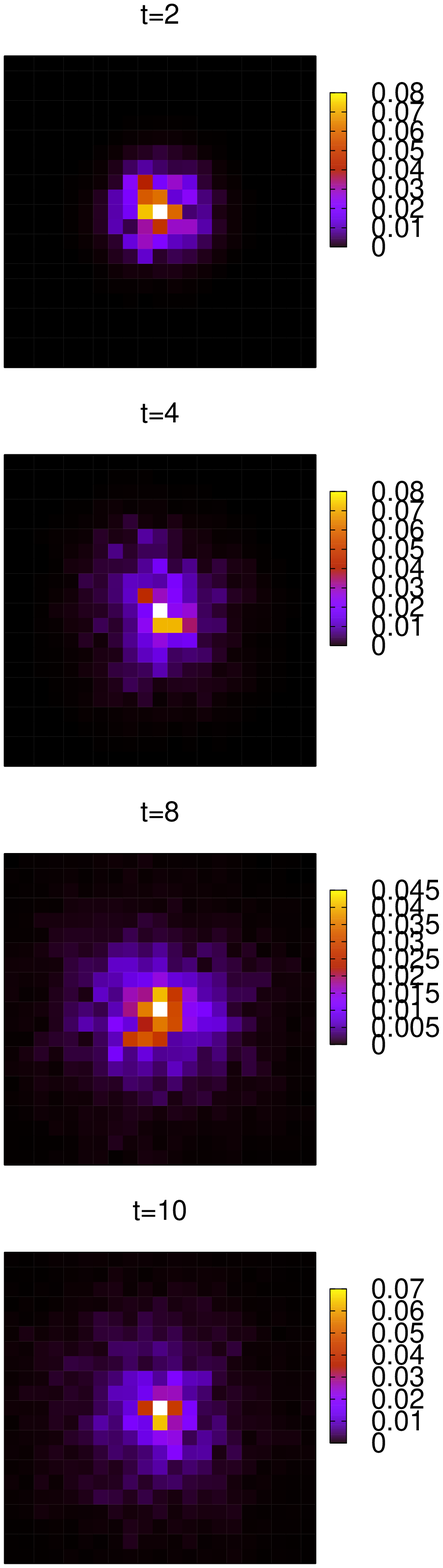}
\end{array}$
\end{center}
\vspace{0in}
\caption{Real time evolution of a wave packet. At low temperature ($T=0$: left panel) it 
propagates due to the checker-board configuration of the heavy atoms. 
At larger temperature ($T=0.2$: right panel), the wave packet is localized due to
a disordered configuration of heavy atoms.}
\label{Fig1}
\end{figure}
\vspace{0in}

{\it localization:} A trapped atomic cloud may be in the state $|i\rangle$, concentrated
at the center of the optical lattice. After switching off the trapping potential the 
dynamics of the atomic cloud is described by the
evolution equation $|\Psi_t\rangle=e^{-iH t}|i\rangle$. The optical lattice remains 
present during the evolution of the
cloud. The expansion of the cloud of light atoms is studied numerically. Depending on the
temperature of the grand-canonical system, we find a spreading of the wave function
at low temperatures but a localized behavior at high temperatures (cf. Fig. \ref{Fig1}).
Apparently, there is a critical regime with some critical temperature $T_c'$, which
separates the spreading behavior from the localized behavior.

The local density of particles at site $r$ of the optical lattice with respect to the 
state $|\Psi_t \rangle$ reads
\[
N_r=\langle\Psi_t|c^\dagger_r c_r|\Psi_t \rangle
=\langle i|e^{iHt}c^\dagger_r c_r e^{-iHt} |i \rangle
\]
with the initial state $|i \rangle=c^\dagger_0 |0 \rangle$. 
The state $|0 \rangle$ is an equilibrium state of the
entire system and can be expanded in terms of energy eigenfunctions and
Boltzmann weights at the inverse temperature $\beta$:
\[
\langle N_r\rangle= {\sum_ke^{-\beta E_k}
\langle E_k|c_0e^{iHt}c^\dagger_r c_re^{-iHt}c^\dagger_0 | E_k\rangle\over
\sum_ke^{-\beta E_k}}
\]
\beq
={1\over Z}{\rm Tr}\left[
e^{-\beta H}c_0e^{iHt}c^\dagger_r c_r e^{-iHt}c^\dagger_0
\right].
\label{dens1}
\eeq 
For the FK model this expression can also be written as a quenched 
average with respect to the distribution of heavy particles 
\cite{ziegler06}:
\beq
\langle N_r\rangle=
\langle {\cal G}_{0r}^\dagger(t){\cal G}_{r0}(t)
\rangle_f 
\label{2pgf}
\eeq
with the single-particle Green's function
\[
{\cal G}_{rr'}(t)=[e^{-ith_c}({\bf 1}+e^{-\beta h_c})^{-1}]_{rr'} \ .
\]
$\langle ...\rangle_f$ is the average with respect to the statistical weight
of Eq. (\ref{distr0}) or Eq. (\ref{idistr}). For a given configuration $\{n_r\}$ 
of heavy atoms
the Green's function can also be expressed by eigenfunctions of the single-particle
Hamiltonian $h_c$ in Eq. (\ref{hamilton2})
($h_c\phi_k=e_k\phi_k$). The spatial properties of these eigenfunctions
determine the spreading of the average density particle density $\langle N_r\rangle$
through the Green's function:
\beq
{\cal G}_{r0}(t)=\sum_k e^{-ie_kt}\frac{\phi_{k,r}^*\phi_{k,0}}{1+e^{-\beta e_k}} \ .
\label{gf0}
\eeq
The denominator represents the Fermi function, since our atoms are fermions.
According to the localization theory, it can be assumed that $|\phi_{k,r}|\sim e^{-|r|/\xi_k}$,
where $\xi_k$ is the localization length.


After a Fourier transformation of the time-dependent density in Eq. (\ref{2pgf}), the
$\omega=0$ Fourier component of $\langle N_r\rangle$ reads
\[
{\bar N}_r=\sum_k\frac{|\phi_{k,r}^*\phi_{k,0}|^2}{(1+e^{-\beta e_k})^2} 
\sim \frac{e^{-2|r|/\xi}}{(1+e^{-\beta e_{k_0}})^2} \ \ \ 
(r\sim\infty) \ ,
\]
where $\xi$ is the largest localization length and $e_{k_0}$ the corresponding
energy level. 
Thus the expansion of the wave packet on large scales is controlled by $\xi$.


The localization length can be studied under the change of length scales by
considering a finite optical lattice of length $L$ and width $M$ \cite{abraham79}. 
In particular, we analyze the change of the localization length with respect to the width $M$.
For this purpose, we define the reduced (or normalized) localization length as
$
\Lambda_{M}=\xi /M
$
and calculate this quantity by means of a numerical 
transfer-matrix approach \cite{mackinnon83}. $\Lambda_M$ either increases (delocalized states) 
or decreases
(localized states) with the width $M$, depending on the system parameters (e.g. the inverse temperature 
$\beta$). There can also be a marginal behavior (e.g. for a special value $\beta_c'$), where 
 $\Lambda_M$ does not change with $M$. The latter indicates the existence
of a phase transition from localized to delocalized states.
A quantitative description of the behavior near $\beta_c'$ can be based
on the one parameter scaling hypothesis \cite{abraham79,mackinnon83}. This states that 
$\ln\Lambda_M$ can be expanded in a vicinity of the critical point $\beta_c'$ as 
\cite{slevin99}
\begin{equation}
\ln\Lambda_M=\ln\Lambda_c\pm A|\beta-\beta_c'|M^{1/\nu} \ .
\end{equation}
For $A>0$ the positive (negative) sign corresponds to delocalized (localized) behavior.
Exponentiation of this equation and using $\zeta=|\beta-\beta_c'|^{-\nu}$
gives
\begin{equation}
\Lambda_M=\Lambda_c \exp\left[\pm A\left(\frac{\zeta}{M}\right)^{-1/\nu}\right]
\equiv g\left( \frac{\zeta}{M}\right) \ ,
\label{scaling}
\end{equation}
where $g$ is the scaling function. Our numerical transfer-matrix approach allows us 
to determine the critical point $\beta_c'$ and the exponent $\nu$, depending on the 
interspecies coupling parameter $U$.



\begin{figure}[!b]
\begin{center}
\includegraphics[width=8cm]{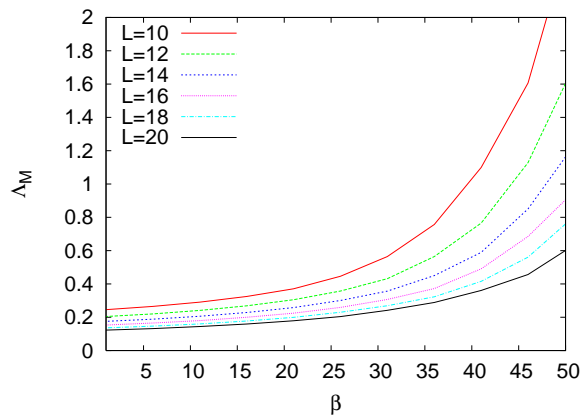}
\end{center}
\vspace{0in}
\caption{The reduced localization length $\Lambda_M$ of light atoms in $d=1$ for interaction strength $U=9$. 
$\Lambda_M$ decreases for increasing system size $M$, which indicates localized states.}
\label{Fig2}
\end{figure}

{\it results:}
First we analyze a one-dimensional system. In this case heavy atoms are always disordered due to
thermal fluctuations. The reduced localization length $\Lambda_{M}$ decreases with increasing
length of the system (cf. Fig. \ref{Fig2}) at any temperature. This indicates that all states are 
localized. On the other hand, the localization length decreases monotoneously with temperature,
as a consequence of the increasing disorder.
Therefore, at sufficiently low temperature the localization length can be 
larger than the size of a finite system. This could be relevant in experiments, where we 
have a finite optical lattice.
 
\vspace{0in}

\begin{figure}[!h]
\begin{center}
\includegraphics[width=8cm]{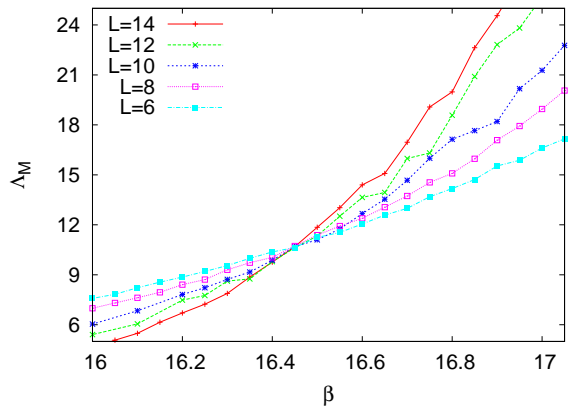}
\end{center}
\vspace{0in}
\caption{Reduced localization length of light atoms for $d=2$ and $U=9$. 
There is a critical inverse temperature $\beta_c'\approx 16.5$, where 
an Anderson transition occurs.}
\label{Fig3}
\end{figure}

\begin{figure}[!h]
\begin{center}
\includegraphics[width=8cm]{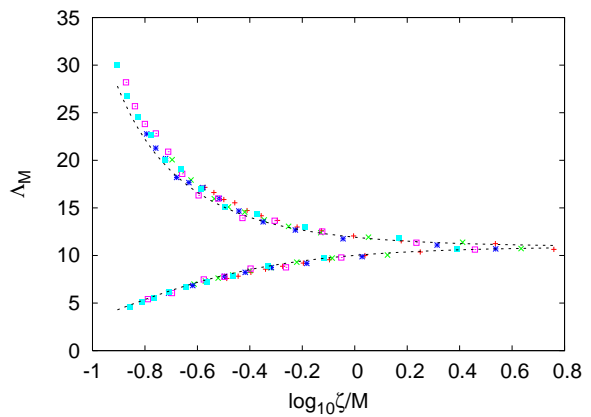}
\end{center}
\vspace{0in}
\caption{One parameter scaling for $d=2$ and $U=9$. The lower (upper) branch represents
(de-) localized states.
The data from the transfer-matrix calculation approach the scaling function of Eq. 
(\ref{scaling}) with $\Lambda_c\approx 10.9$, $A\approx 0.09$ and $\nu\approx 0.88$
(dashed curves).}
\label{Fig4}
\end{figure}

In two dimensions the behavior is more complex. First of all, the heavy atoms can form an ordered state
at low temperatures and a disordered state at high temperatures \cite{ates05,maska08}. As long as $T>0$, 
thermal excitations in the ordered state lead to correlated fluctuations of heavy atoms. There is a
second-order phase (Ising) transition  with a divergent 
correlation length at the critical temperature $T_c$. 
The corresponding distribution of heavy atoms provides a complex random environment for the light atoms. 
Our numerical transfer-matrix approach
finds a transition from localized states at high temperatures to delocalized states at low temperatures.
There is a critical temperature $T_c'$, where this transition takes place. 
For instance, at low temperatures and half filling (i.e. for $\mu_f=\mu_c=U/2$), the heavy atoms 
are arranged in a staggered configuration with weak thermal fluctuations. Using the approximated 
distribution of Eq. (\ref{idistr}), the effective spin-spin coupling ${\bar t}_c^2/2U$ leads to the
critical temperature $T_c\propto {\bar t}_c^2/2U$.
The result for the reduced localization length at $U=9$ (measured in units of ${\bar t}_c$) 
is shown in Fig. \ref{Fig3}. All curves cross at $\beta_c'\approx 16.5$, 
indicating a localization transition. With these parameters the Ising transition
is at $\beta_c\approx 15.9$. Therefore, the localization transition occurs in the ordered phase 
of the heavy atoms. 
The one-parameter scaling function of Eq. (\ref{scaling}) with
\beq
\Lambda_c\approx 10.9 ,\ \  A\approx 0.09 , \ \  \nu\approx 0.88
\label{parameters}
\eeq
fits the data of the transfer-matrix calculation (cf. Fig. \ref{Fig4}).

\vspace{0in}


In conclusion, we have discussed a mixture of two fermionic species with different masses
in an optical lattice, using the Falicov-Kimball model. The heavy atoms are represented as 
Ising spins and the light atoms as quantum particles. The latter tunnel in a random environment
which is provided by a correlated distribution of heavy atoms. The distribution of the heavy
atoms is given by an Ising-type model, which undergoes a second-order phase transition in $d=2$
from staggered order to disorder. Depending on the
dimensionality ($d=1,2$) of the atomic system and the physical parameters (e.g. temperature or interaction
strength), the quantum states 
of the light atoms are either localized or delocalized. All states of light atoms in a one-dimensional 
fermionic mixture are localized. In a two-dimensional mixture these states are localized at high 
temperatures and delocalized at low temperatures. 
Such a system can be realized experimentally as a mixture with two spin-polarized fermionic species.

\bibliography{reference}

\end{document}